# Ultrahigh measured electrostrain in polycrystalline bulk piezoceramics: Role of mechanically relaxed grains

Gobinda Das Adhikary and Rajeev Ranjan

Department of Materials Engineering, Indian Institute of Science, Bangalore-560012, India.

**Abstract**

Recently we found that when conventional bulk polycrystalline piezoceramics discs (~ 10 mm diameter) are thinned down to thickness ~ 200 microns or below, the measured unipolar electrostrain (UES) values increased phenomenally (https://arxiv.org/abs/2208.07134). Here we demonstrate that this anomalous phenomenon correlates strongly with a substantial decrease in the degree of mutual mechanical clamping between the grains which in turn enables significantly enhanced switching of the ferroelastic domains.

rajeev@iisc.ac.in

-------------------------------------------------------

Piezoelectric materials change shape on application of electric-field (converse effect) and develop voltage on application of force (direct effect) [1]. Both these aspects are used in a variety of technological applications. Recently we found that when the thickness of conventional piezoceramic discs (10 mm diameter) is reduced the measured electrostrain increased significantly [2]. For example, the measured electrostrain values of the morphotropic phase boundary compositions of PZT increased from ~ 0.3 % to 2.5 % when the thickness was reduced from 700 microns to 200 microns [2]. A similar behavior was found for the morphotropic phase boundary compositions of Sn-modified $BaTiO_3$ and a $Na_{0.5}Bi_{0.5}TiO_3$-based piezoceramic. This phenomenon appears to be common across piezoceramics. Field driven strain measurements were carried out with a MTI Fotonic displacement sensor and sample holder assembly supplied by Radiant Technologies Inc, USA. The upper and bottom electrodes of the sample holder was ~ 6 mm diameter. During the measurements, the center of the circular disc shaped pellets (10 mm diameter) was kept almost at the center of the electrodes of the sample holder. The details can be found in ref [1].

To gain better structure and microstructural insights into this anomalous phenomenon, we carried out x-ray diffraction measurements as a function of thickness of the disc. Experiments were carried out using laboratory rotating anode x-ray (Cu-K$\alpha_1$ radiation) diffractometer (9 kW, Rigaku Smart Lab model). A Johansen monochromator in the incident ensured only K$\alpha_1$ radiation falling on the specimen. The diffraction measurements were performed in reflection geometry. For the sake of simplicity and ease of interpretation, here we focus our attention on a tetragonal composition of PZT, PbZr$_{0.45}$Ti$_{0.55}$O$_3$ (PZT45), far away from the MPB.

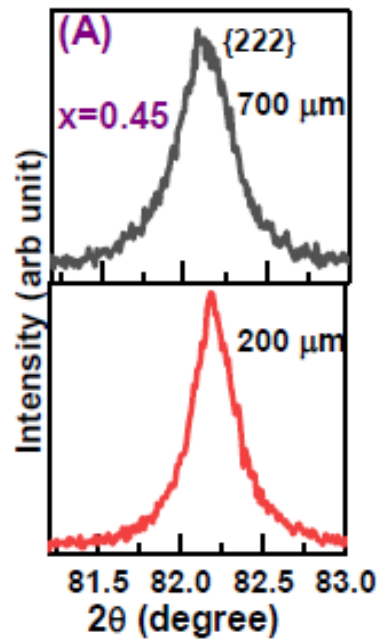

Fig. 1. X-ray diffraction profiles of {222} Bragg reflection of PZT45 collected from 700 micron and 200-micron discs.

The diffraction patterns (not shown here) of both 700- and 200-micron discs look similar. A careful examination, however, revealed that the width of {222} reflection obtained from the 200-micron disc is smaller than that obtained from the 700-micron disc pellet, Fig. 1. We may note that diffraction patterns were obtained from the same disc specimen by reducing the thickness. A systematic variation of the full width at half maximum (FWHM) of {222} profile is shown in Fig. 2.

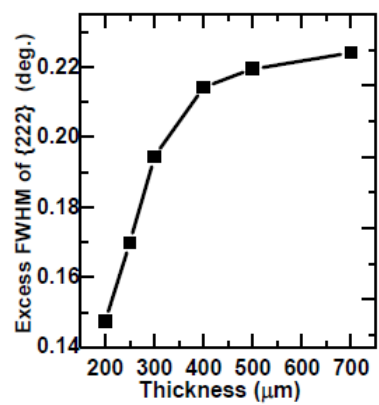

Fig. 2. Variation of excess FWHM of {222} profile of PZT45 with thickness variation.

We may note that although the diffraction patterns were obtained from the grains near the surface regions of the discs ( < 10 microns beneath the surface of the disc), the systematic change in the width of the Bragg profile confirms a direct influence of the bulk grains (deeper than 10 microns) on the surface grains from which we extract the information. Since the average grain size (~ 3- 5 microns ) is sufficiently large, and same across the disc volume, the change in the width of the Bragg profiles can be attributed to change in the degree of the strain inhomogeneity on the grain scale [3]. In the present case, the strain heterogeneity can be attributed to the mutual clamping of the grains. The mutual mechanical clamping of the grains tend to suppress the change in the shape of the grains as the piezoceramic cools through the Curie point and introduce inhomogeneous stress [4]. We estimated the inhomogeneous strain/microstrain from the FWHM of the {222} profile using the expression

$$\varepsilon(\%) = \frac{\beta}{4\tan\theta} * 100,$$

where $\beta$ is FWHM of Bragg peak, and $\theta$ is Bragg angle. In Fig 2, $\beta$ is represented by the excess FWHM. By excess FWHM we mean the width of the Bragg profile which is more than the width measured on powdered specimen of the same composition (obtained by breaking the pellet to powder). The grains of a powdered specimen are free and there is no mutual clamping. In this definition, the excess FWHM (in Fig 2) is zero for unclamped grains. The non-zero values of the excess FWHM can be taken as a measure of the degree of mutual mechanical clamping the grains experience in the ceramic body.

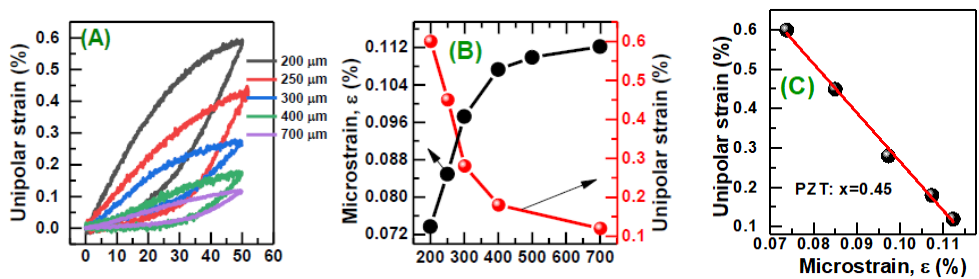

Fig. 3: (A) Unipolar electrostrain (at 1 Hz) of PZT45 measured on discs of different thicknesses. (B) Variation of the thickness dependence of unipolar electrostrain (at 50 kV/cm) and microstrain. (C) Variation of the measured macroscopic electrostrain with the microstrain/strain inhomogeneity.

Fig. 3 shows the thickness dependence of the unipolar electrostrain of PZT45. The unipolar electrostrain (UES) on the 700-micron disc PZT45 measured at 50 kV/cm is ~ 0.12 % (Fig. 3A). When the thickness is reduced to 200 microns, the UES increased ~ 0.6 %, i.e. by almost five fold. Most importantly, there is a one-to-one correspondence between the decrease in the microstrain and increase in the electrostrain, Fig. 3B. In fact, the unipolar strain follows a linear increasing relationship with decreasing microstrain/strain heterogeneity in the grains, Fig. 3C. This correlation confirms that the considerable increase in the measured electrostrain below 400 micron is a direct consequence of the weakening in the degree of mutual mechanical clamping between the grains as the discs is thinned down.

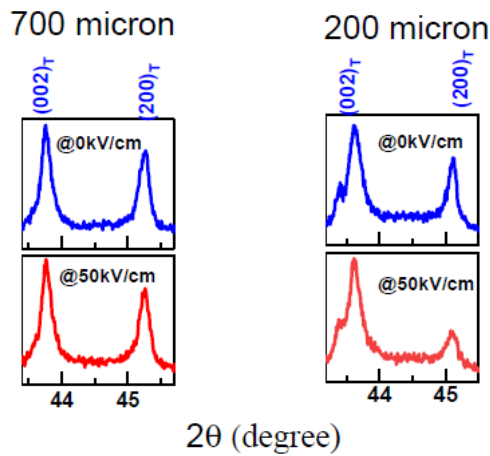

Fig. 4: Change in the relative intensities of 002 and 200 tetragonal peaks on application of electric field (50 kV/cm) on 700 micron and 200 micron PZT45 discs. The small peak on the left of 002 in the right panel is from the electrode material.

Fig 4 shows the change in the relative intensity of the Bragg profiles of the 002 and 200 tetragonal Bragg peaks of PZT45 on application of field (50 kV/cm). The data shown in Fig. 4 are from the second cycle. The first cycle poles the specimen. The changes in the pattern brought about by the field in second cycle are reversible in nature and should be correlated with the reversible electrostrain shown in Fig. 3(A). The higher intensity of the 002 peak as compared to the 200 peak at 0 kV/cm is due to the texturing introduced during the poling (first) cycle. On application of 50 kV/cm the extent of the texture (as measured in terms of the intensity ratio $I_{002}/I_{200}$) increased further. Most importantly, this ratio is substantially increased in the 200-micron thick disc as compared to the 700-micron disc of PZT45. The fraction of reversible switching of the ferroelastic domains estimated from the intensity ratios was found to be nearly five time higher in the 200-micron disc than in the 700-micron disc. This corelates well with the fivefold increase in the electrostrain in the 200-micron disc of PZT45, Fig 3.

In summary, we demonstrate that the significant increase in the measured electrostrain in thin piezoceramic discs correlates directly with the decrease in the degree of mutual mechanical clamping between the grains of the piezoceramic disc. A direct consequence of the reduction in the degree of mutual mechanical clamping is a considerable increase the switching of the ferroelectric-ferroelastic domains which in turn cause large electrostrain. Though not given due attention by the research community, we argue that the degree of mechanical relaxation of the grains in piezoceramics is an important factor in the large electrostrain reported in different piezoceramics from time to time.

**Acknowledgements:** RR acknowledges the fundings received from Science and Engineering Research Board (SERB), Indian Institute of Science (IISc), Naval Research Board (NRB) India, for conducting research on piezoceramics over the years.